# Heterogeneously Integrated Diamond–on-Lithium Niobate Quantum Photonic Platform

Sophie W. Ding[1,†], Chang Jin[1], Zixi Li[2], Nicholas Achuthan[1], Kazuhiro Kuruma[1,3], Xinghan Guo[2], Brandon Grinkemeyer[4], David D. Awschalom[2,5], Nazar Delegan[5], F. Joseph Heremans[2,5], Alexander A. High[2,5,*], and Marko Loncar[1,*]

[1]John A. Paulson School of Engineering and Applied Sciences, Harvard University, Cambridge, Massachusetts 02138, USA.
[2]Pritzker School of Molecular Engineering, University of Chicago, Chicago, IL 60637, USA.
[3]Research Center for Advanced Science and Technology, The University of Tokyo, Meguro-ku, Tokyo 153-8505, Japan.
[4]Department of Physics, Harvard University, Cambridge, Massachusetts 02138, USA
[5]Q-NEXT, Argonne National Laboratory, Lemont, IL 60439, USA.
*E-mail: ahigh@uchicago.edu; loncar@g.harvard.edu

## Abstract
Diamond photonics has enabled efficient, high-fidelity interfaces for diamond quantum memories and is predicted to be a critical component of modular quantum networks. However, scalable network architectures require spatial, temporal, and spectral control of photons emitted by or directed towards quantum memories, which, in turn, relies on nonlinear and electro-optic functionalities that diamond alone cannot provide. Here, we demonstrate heterogeneous integration of a thin-film lithium niobate (TFLN) platform, featuring strong chi-2 nonlinearity and electro-optic effects, with thin diamond films. We demonstrate high-Q diamond photonic crystal cavities (Q factors exceeding $5 \times 10^4$ at 735 nm) that are lithographically aligned with TFLN photonic backbone and critically coupled to it. This allows us to realize low-loss diamond-TFLN "escalators" (loss ~1 dB/coupler) that support efficient light transfer between the diamond and TFLN layers. By operating our platform at cryogenic temperatures (5K), we demonstrate the collection of photons emitted from silicon vacancies (SiVs) embedded within a diamond structure via the TFLN photonic circuit. By combining thin-film diamond with TFLN, this approach establishes a scalable route toward integrated photonic circuits for practical quantum networking and other technologies.

## Introduction
Diamond color centers, with their optically addressable and long-lived spin qubits, have emerged as leading candidates for scalable quantum networks[1–3]. Recent demonstrations of multimode entanglement and long-distance links using nitrogen- and silicon-vacancy centers highlight the platform's potential[4,5]. Yet a practical quantum network requires more than just spin qubits: it also requires switches, modulators, frequency converters, and other photonic components. Diamond itself lacks the nonlinear and electro-optic (EO) functionalities needed for such operations. Thus, progress toward large-scale, high-performance networks depends not only on advances in diamond devices but also on their integration with complementary platforms via low-loss interfaces.

Thin-film lithium niobate (TFLN), with exceptionally low optical loss, strong optical nonlinearity, and large-scale manufacturability, has recently emerged as a leading platform for high-performance nonlinear photonics and opto-electronics[6]. TFLN also offers the potential for heterogeneous integration with other photonic materials, such as SiN[7,8] and diamond [9–11].

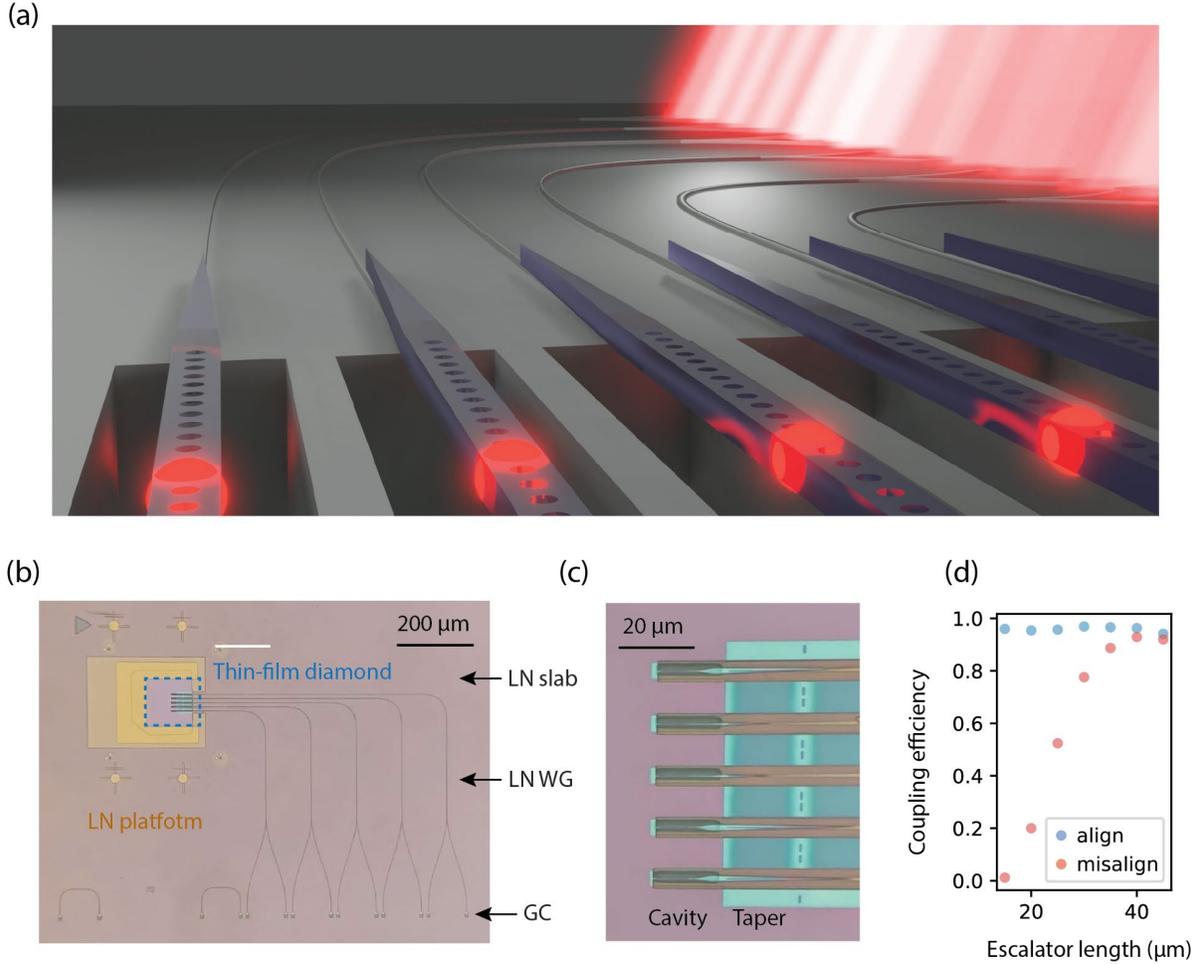

Fig. 1. (a) An illustration of the LN-diamond photonic platform. Photons emitted by the SiV are captured by free-standing high-Q diamond photonic crystal cavities or waveguides, transferred to LN via high-efficiency LN-diamond double taper escalators, and coupled out through the LN grating coupler centered around 740 nm.  (b) A microscope image of the fabricated devices. The light pink region is the LN "slab", and the darker lines are LN rib waveguides. The dashed square indicates the region where a diamond thin film was transferred onto the LN platform. The diamond is secured by metal "stickers", as seen in yellow. (c) A zoomed-in image of the diamond photonic crystal cavities and the diamond-LN escalators of different lengths. The cavities are suspended over a trench. (d) Simulated coupling efficiency of the escalator with long diamond taper, assuming both perfect alignment and 100-nm lateral
 misalignment. When there is no misalignment, the coupling efficiency is close to 96±1%, but with the misalignment, the efficiency drops quickly as the taper shortens, from 93% at 35 µm to 1% at 10 µm.

Previous approaches of integrating diamonds with other photonic platforms have relied on several distinct transfer and bonding methods. In the pick-and-place method [11–15], diamond devices, including photonic crystal cavities and waveguide arrays, are first fabricated on diamond and then lifted out and transferred onto another photonic platform. The transfer printing method [16,17], relies on a stamp to pick up a diamond film or fabricated structure and press it onto another photonic platform. Finally, the flip-chip bonding method [18,19] uses a die bonder to bond a diamond film to another flat substrate using metal or other bonding agents. Depending on the chosen transfer method or the transferred structure, bonding agents may be required. Pick-and-place and flip-chip bonding usually require some "glue", which can cause losses or luminescence at short wavelengths. Transfer printing can achieve direct bonding under certain conditions [17]. A low-loss bonding agent or no bonding agent is preferred for minimizing loss. The achieved platform typically features either evanescent [15,16] or direct coupling between the diamond emitters and light, depending on whether the light is guided mainly within the diamond or in the surrounding material. The latter tends to feature much stronger coupling and better photon collection efficiency from the emitter, and is preferred.

TFLN devices often rely on a rib waveguide mode that features a thin layer of LN ("slab") that is not etched through and on top of which metal electrodes are deposited to ensure efficient EO control. This "slab", combined with non-vertical sidewalls typical in etched TFLN waveguides, makes it difficult to efficiently ($\sim < 1$ dB at 740 nm, e.g.) couple light out of the chip, through grating couplers, edge couplers, or taper fiber couplers [20]. This limitation further motivates the development of direct high-efficiency coupling between TFLN and other materials.

When integrating TFLN and diamond, several key aspects need to be considered to evaluate the viability of the approach. (1) Scalability, reproducibility, and yield: the process should ideally allow for the parallel fabrication of devices with consistent performances, and have a robust fabrication process that can be repeated reliably. (2) Preservation of material and device quality: the integration should preserve high optical Q and operating wavelength of the diamond cavities, as well as optical and spin coherence of embedded color centers. It should also preserve the TFLN circuit's integrity. (3) High coupling efficiency and mode matching between the diamond nanophotonic cavity and the LN waveguide: this requires engineering of the tapers at the interfaces and near perfect alignment, which is essential for quantum applications.

We present a heterogeneous integration method that unites state-of-the-art diamond photonic crystal (PhC) cavities with TFLN circuits using transfer printing of thin-film diamond and lithographically defined alignment of photonic structures. In our approach, the TFLN optical circuit, consisting of optical waveguides, Y-splitters, and grating couplers (GCs) all operating around 740 nm, is first fabricated. On one side, TFLN waveguides feature tapers sandwiched between unetched TFLN regions for diamond mounting. Next, a thin-film diamond is transferred onto these mounting areas using the direct bond method [17]. Diamond photonic crystal cavities are then fabricated using a combination of electron beam lithography (EBL) and reactive ion etching (RIE). Importantly, the EBL-written diamond beams with the coupling tapers are aligned to the underlying TFLN circuit, where we expect very precise alignment between diamond and TFLN coupling tapers. The intended or technical limit for alignment error is ~25 nm, imposed by the EBL tool used (Elionix ELS-7000; quoted overlay accuracy is 25 nm), thereby allowing us to make efficient TFLN-diamond escalators. As shown in Fig. 1 (a), the final array of fabricated diamond PhC cavities is suspended over the trenches, which are etched into the TFLN and SiO2. The light emitted by silicon vacancies (SiVs) embedded in the diamond structure travels through the escalator to the TFLN chip, and eventually couples out through the grating couplers after the Y-splitters. Our approach also enables parallel fabrication of device arrays, limited primarily by the size of the thin-film diamond. The latter is expected to increase in size as the growth and transfer technologies mature[19].

In this work, we fabricated five diamond PhC cavities that all exhibit on-target resonances (average 0.33% error) and have high quality factors (Qs), consistent with previous state-of-the-art results on the diamond platform[21]. The critically coupled cavity at 735 nm achieves a Q of $5.3 \times 10^4$ (scattering Q of $\sim 1.1 \times 10^5$). The diamond-LN escalator efficiency is measured to be up to 91 % (-0.42 dB/escalator) at 775 nm. We characterized our platform at cryogenic temperatures (5K) and observed ensemble SiV emission via the LN grating couplers, as well as the high-Q cavity mode reflection dip. By combining diamond and TFLN with low loss, this platform offers a route to integrate quantum materials with nonlinear and EO functionalities toward integrated photonic circuits for quantum networking and beyond.

## Device design and concept

To preserve the performance of both visible TFLN and diamond devices, we decided to design escalators around material stacks used in state-of-the-art TFLN modulators at visible wavelengths[22] and high-Q diamond cavities[21]. As shown in Fig. 1(b), we fabricated 5 sets of devices. Within each set, there is a diamond PhC cavity on the left end, as shown in Fig. 1(c). To the right, there is an escalator formed by the diamond and LN tapering in width, followed by a segment of LN single-mode waveguide that leads to a Y-splitter, and two GCs. All the components are designed for wavelengths around 737 nm, the zero-phonon line (ZPL) of the SiV.

The diamond PhC cavities are suspended and single-sided: coupled to the waveguide on one side only. This reflection-type PhC cavity design is the workhorse of SiV-based quantum memory nodes[1]. The cavities are designed using FDTD EM simulations, with the same design parameters as in our previous work[21]: the width and thickness of the beam are 370 nm and 160 nm; the hole radius is 65 nm; the lattice constant is varied to target the fundamental TE modes at 730, 740 and 740 nm; for 740nm, the unit cell length is 255 nm.

The LN waveguide, grating couplers, and Y-splitters are designed using FDTD EM simulations (Tidy3D) for wavelengths around 740 nm. The X-cut LN layer is 200 nm thick on 3 μm of SiO2, which sits on a Si handle wafer. Since LN etch has a characteristic sidewall angle of 60°, it is accounted for in the simulation and subsequent design. The waveguide width (330 nm top width) and thickness (100 nm etch depth and 100 nm "slab" thickness) are chosen to ensure single-mode operation and avoid TE/TM coupling that is common at LN waveguide bends due to the high birefringence of the material. This allows us to maintain the polarization of the light as it propagates on-chip, which is critical to avoid polarization-dependent loss at the GC/fiber interface and to ensure proper operation of our diamond PhC cavities. The latter has TE band gaps (see SI) only and thus can couple to or reflect only TE-polarized light for detection. We note that, in our design, any residual TE/TM coupling reduces escalator coupling efficiency and contributes to total photon loss. Finally, this geometry is consistent with visible-wavelength TFLN electro-optic modulator platforms[22], and therefore compatible with electrode integration for active control.

The GCs feature a 465nm period with a 31% duty cycle and are designed to achieve maximum coupling efficiency around 740 nm for an angle-polished (-8°) fiber. The GCs are spaced 127 μm apart horizontally (Fig. 1(b)) to be compatible with a fiber array used to couple light in and out of the chip. The escalator efficiency is measured using the cutback method, which removes the losses related to the GC. The circuit is intentionally designed to separate the excitation and collection ports to avoid back-reflection noise from the chip bottom during characterization. To accomplish this, we use adiabatically tapered Y-splitters that ensure equal splitting between the two arms with minimal excess losses (see SI).

The escalator is achieved through an adiabatic LN-diamond taper. While the refractive indices of the two materials are similar (~2.2 vs. 2.4), which should simplify the design, the situation is complicated by the presence of the unetched LN slab in the LN rib-waveguide. The LN tapers are formed by linearly reducing the LN waveguide top-width from 330 nm to 40 nm over a 40 μm length. On the diamond side, we adopted spear-tip shaped taper designs (Fig. 2(a)) and considered two different lengths: long and short, with lengths of 35 μm and 11 μm, respectively. In the case of long (short) tapers diamond waveguide width is first increased from 60 nm to 1.66 μm (860 nm) and then reduced to 370 nm. Based on numerical modeling, both designs can achieve > 95% coupling efficiency when the alignment is perfect at 770 nm (see SI for other wavelengths). The spear-tip diamond coupler is also chosen to increase misalignment tolerance for a given overlap length. When misaligned, the long taper has a higher coupling efficiency, indicating greater misalignment tolerance. Fig. 1(d) shows the simulated effect of 100 nm misalignment on tapers of different lengths, which have different overlaps with the LN taper. The long taper design is illustrated in Figs. 2(b) and (c) through the mode profile of the field propagating between the diamond and the LN tapers. The short taper design is illustrated in Fig. 2(d) and (e), where high efficiency can also be achieved with good alignment. The purpose of testing both is to understand the trade-off between alignment tolerance and the extra scattering loss introduced in a long taper region using this method.

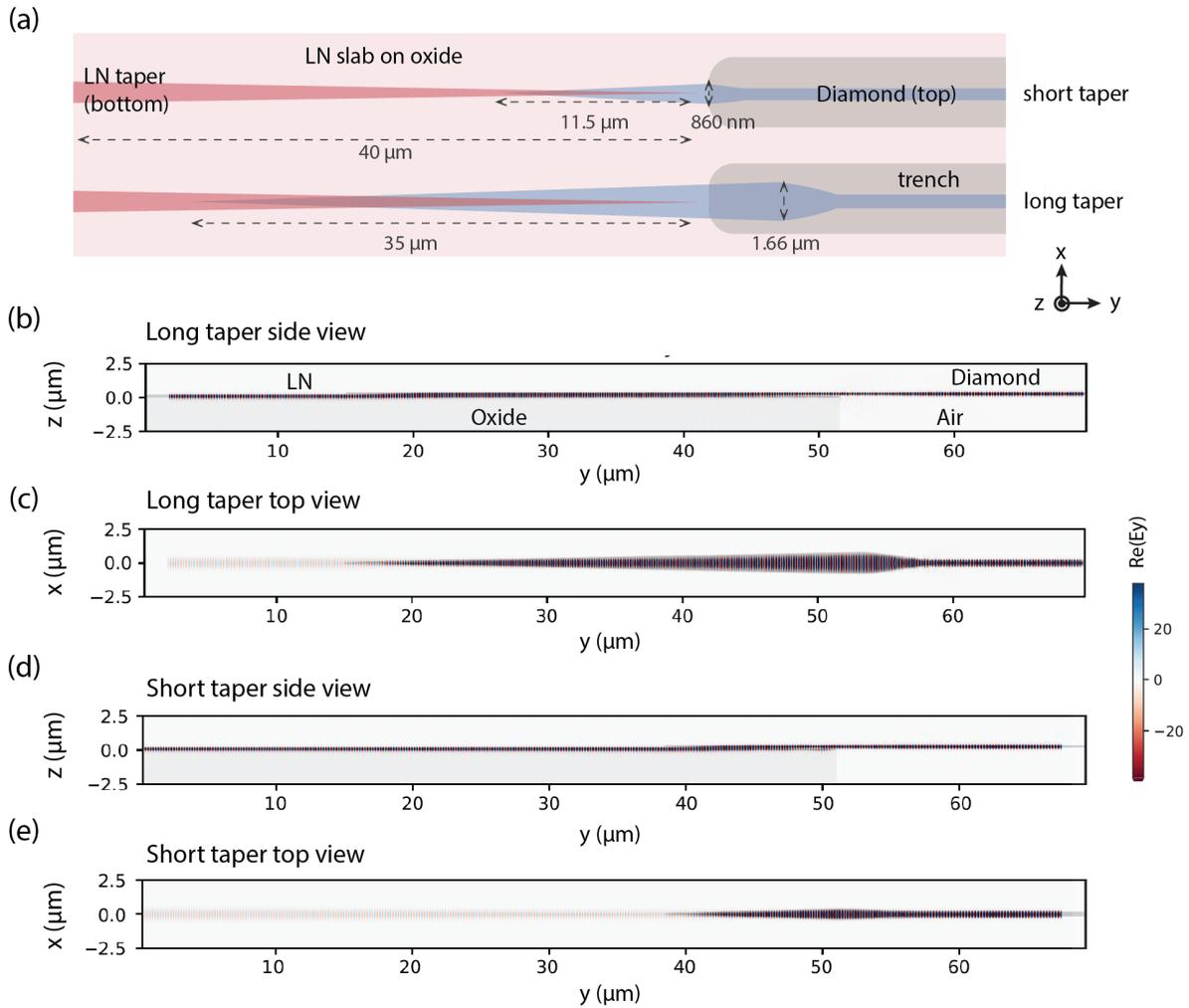

Fig. 2 Heterogeneous integration of LN and diamond: (a) The geometry schematic for the LN and diamond taper that make up the escalator, in both the long and short taper configurations. (b-e) The field profiles of light propagating through the LN-diamond interface, where it adiabatically transforms between the LN and the diamond waveguide modes, with both long and short diamond tapers. The long taper (b,c) is designed to tolerate more misalignment, and the short taper (d,e) is designed to mitigate any fabrication-induced loss in the long taper.

## Device fabrication

Device fabrication begins with the TFLN backbone circuit. The chip starts with 200 nm LN on 3 μm of oxide on a silicon substrate (Fig. 3 (a)(1)). First, the LN circuit is defined by EBL using MAN e-beam resist, followed by inductively coupled plasma (ICP) RIE using Ar chemistry. The etch depth is 100 nm, leaving behind a 100 nm TFLN slab (Fig. 3 (a)(2)). Next, the trenches are defined using SPR220-3.0 and optical lithography, aligned to previously EBL-defined structures. RIE with Ar and $C_3F_8$ are subsequently used to etch through the LN and > 1μm of SiO2 (Fig. 3 (a)(3)).

A 200 μm x 200 μm thin film diamond (160 nm thick) is transferred onto the platform through a transfer printing process with a minimal amount of alignment requirement (Fig. 3 (a)(4))[17]. The samples are cleaned with acetone after transfer to remove residual contaminants from the transfer steps. To secure the bond while maintaining the integrity of the TFLN, the sample is annealed at 400 °C in air (see SI).

As the first step in fabricating the heterogeneous chip, a 200 nm Cr/Au layer is deposited to secure the membrane, as shown in Fig. 3(a)(5). In future applications, electrodes for LN modulators can also be defined at this step without interfering with subsequent steps. Then, the diamond fabrication follows a known process[21], where the cavity is defined by EBL and RIE etching of O2, as shown in Fig. 3 (a)(6) and (7). BOE 7:1 is used to remove the SiN mask as the final step, further undercutting the oxide underneath the diamond beams. The resulting chip features suspended diamond PhC cavities adiabatically tapered into a LN photonic circuit with a 100 nm slab (Fig. 3(a)(8)).

The fabricated devices are shown in Figs. 3(b)-(g). The components on LN appear smooth, with no visible defects (Fig. 3(b)). The diamond cavities are indeed well aligned with the LN taper over the trench in both the long- (Fig. 3(c)) and short-taper (Fig. 3(e)) designs. The alignment error is of the same order as the measurement error in the SEM images, consistent with the nominal EBL overlay error of < 30 nm. Detailed examination of the diamond taper (Fig. 3(d)), cavity (Fig. 3(f)), and trench (Fig. 3(g)) shows no obvious fabrication imperfection or deviation from design that is not accounted for. There are a few minor but noticeable defects, however (SEM images see SI). At the edge of the thin-film diamond, there is a thin line (~50 nm), which is likely the residue of the transform printing agent that was not fully cleaned off. The O2 etch of diamond also introduces a shallow dent (~< 10 nm) in the LN that was exposed during the write step. These defects would contribute to scattering losses, which will be included in measurements of the escalator coupling efficiency. Another defect is on device 5, which shows an unexpected instance of misalignment: in this case, the wet process (BOE) dislodged the written and well-aligned taper at the tip, and slightly tilted it, so that there is a slight rotation or bending of the diamond taper, centered somewhere on the diamond waveguide.

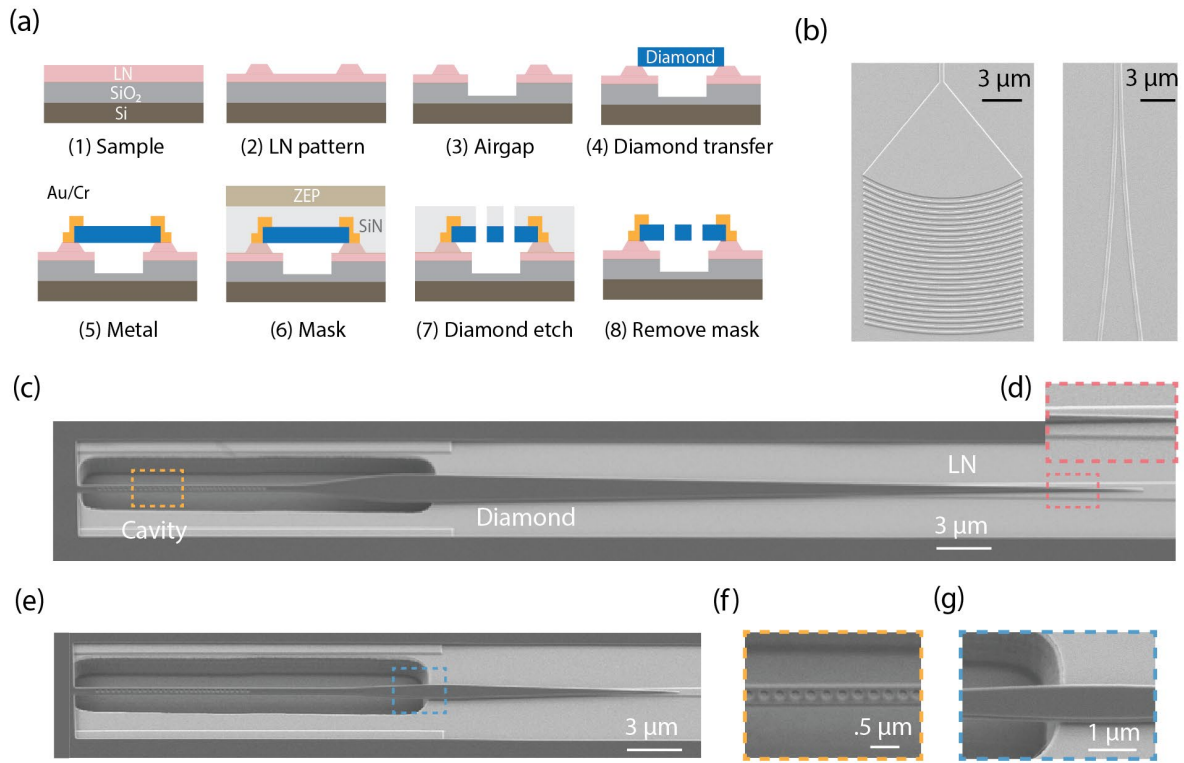

Fig 3. The fabrication of LN and diamond photonic crystal cavity: (a) The diagram of the fabrication flow. (b) The fabricated LN GC (left) and Y-splitter (right). The aliasing in the GC image is a display artifact, and not on the actual device. The LN etch depth is 100 nm on a 100 nm stack. There is no visible roughness on the waveguide. (c)(e) Two fabricated diamond-cavity-LN interfaces using different taper designs. The SEM images are taken at a 30-degree tilt. The diamond tip looks almost transparent (white) to electrons since it's very thin. It might give the illusion that the tip is misaligned, but zooming into the image reveals that both tapers are well aligned. (d) A zoom-in image of the diamond (dark) and LN (light) taper overlap. The amount of misalignment is minimal. (f) A zoom-in image of the diamond photonic crystal cavity suspended over an LN-SiO2 trench. (g) A zoom-in image of the diamond mode converter over the LN edge. There is no visible roughness on the diamond cavity or waveguide

## Escalator coupling efficiency and cavity characterization

The coupling efficiency over different wavelengths is characterized by sending supercontinuum white light through one GC and measuring the light reflected from the diamond cavity with the other GC. More specifically, the data are taken by a fiber array over the grating couplers with an OSA (optical spectrum analyzer), and the fiber array moves between devices and realigns itself automatically and locally between each wavelength sweep.

The diamond PhC is designed to have a TE bandgap of 700-800 nm across all devices, which functions as a reflector at the end of the circuit. We can extract the escalator's coupling efficiency by measuring the light intensity before and after the GCs, and subtracting the GC, Y-splitter, and propagation losses. In other words, the total efficiency as a function of wavelength, $\eta_{tot}$ ($\lambda$) is:

$$\eta_{tot} = I_{in}/I_{out} = (\eta_{escalator} \times \eta_{y-splitter} \times \eta_{prop})^2 \times \eta_{GC,tot}$$

The losses are counted twice because the light is reflected back and encounters the interfaces or losses twice. The $\eta_{escalator}$ accounts for both the photon transfer efficiency and mode matching between the diamond PhC and the LN waveguide. The circuit is designed to separate the excitation and collection ports to avoid noise from back reflection, so the Y-splitter is used. It introduces a simulated $\eta_{y-splitter}$ of 3.1 dB per pass in this wavelength range (see SI). The total GC loss, $\eta_{GC,tot}$, can be calibrated out using the cutback method, in which the short GC loop's spectra are collected and subtracted from the measurements. The propagation loss can be calculated from measurements of reference devices of different lengths fabricated on the same chip. The LN propagation loss $\eta_{LN}$ is extracted from waveguide measurements at the relevant wavelengths, giving a loss of 0.3 dB/mm (see SI).

The measured coupling efficiency, $\eta_{escalator}$+ $\eta_{LN}$ in the 680-800 nm range, is shown in Fig. 4(a) for the 5 sets of devices (raw data in SI). They are averages of two consecutive scans to account for drift in the setup's power and optical alignment. Since the GC cut-back calibration method and measurement setup introduce oscillatory artifacts in the measured spectra, we quote the coupling efficiency averaged over a modest range of wavelength window as a more representative and meaningful figure of merit.

We note there is a trend of increasing coupling efficiency with longer wavelengths. This points to scattering-related loss mechanisms in the circuit or the escalator. In modeling, this escalator design shows either very little wavelength sensitivity (long taper or aligned short taper) or the opposite trend (misaligned short taper) in this wavelength range, as discussed in the SI. In the measurements of the LN waveguide propagation loss, the dependence on wavelength is also not as strong, and therefore is not a dominating factor (more details in SI). For loss α, in dB/length, there are a few possible loss channels in this case: (1) Rayleigh scattering: if the loss is dominated by very small particles or defects, the loss is α~$1/\lambda^4$; (2) sidewall roughness (Payne-Lacey): the roughness on the sidewall contributes to loss, inversely related to the wavelength, with some assumptions usually leads to scalings like α~ $1/\lambda^2$ [23]. Therefore, a quadratic interpolation based on all measured curves is overlaid on the curves in Fig. 4(a) as a visual guide.

On our chip, devices 3 and 4 have short tapers, and devices 1, 2, and 5 have long tapers. The fact that devices 1, 2, and 4 exhibit comparable reflection spectra features and higher coupling efficiencies suggests that, when misalignment is not a dominant loss mechanism, short and long tapers are both good, consistent with the simulation results. Device 3 has a flatter spectrum and higher loss. This could indicate a small degree of misalignment of the short taper, in addition to the previously identified loss mechanisms. We did not observe obvious misalignment under SEM (<~50 nm), but the shorter design is sensitive to even small misalignments and could have accounted for it. Device 5 is the worst-performing long taper device. We have indeed observed some tilting of this taper and debris on it (SiN film from the mask).

At around the peak, which is 775 nm (760 to 790 nm), both the "raw" coupling efficiency on chip $\eta_{escalator}$+ $\eta_{LN}$ and the "corrected" one, $\eta_{escalator}$ of the LN-diamond tapers are extracted, as shown in the inset of Fig.4 (a). The best device shows $\eta_{escalator}$ up to -0.42 dB for device 1, which has a long taper. The best short taper has $\eta_{escalator}$ = -1.26 dB. This means longer tapers are more robust and less sensitive to fabrication imperfections, and tend to perform better.

After examining the escalator properties, we also characterized the diamond cavities at room temperature. The cavity spectra are obtained by collecting photoluminescence from SiV and other unintentional defects (e.g., NVs) in the diamond when the cavities are illuminated with out-of-plane green excitation, similar to the method used in our previous works[21,24]. The spectrum is shown in Fig. 4(b) for the cavity designed to target 737 nm. The measured resonances of all five cavities, designed for 730, 740 (3 devices), and 750 nm, respectively, are extracted and closely match the designed resonances, as shown in the inset of Fig. 4(b). The average error is 0.33%, consistent with thin-film diamond fabrication without the TFLN backbone. The cavities show high Qs (>20k), except for cavities not designed to be very overcoupled, and are limited by the spectrometer's resolution. To fully resolve the resonance features of the diamond photonic crystal cavities and to verify that the heterogeneous device is compatible with cryogenic operation, we proceed to measure the chip at 5 K.

## SiV characterization through the circuit (5K)

For the cryogenic optical setup at 5K, we used a 100x objective with a high NA of 0.95 to observe the SiVs confocally, and used a 20x objective with a moderate NA of 0.4 to achieve good coupling to the GCs and provide a generous field of view. The experimental setup for both configurations is illustrated in Fig. 4(c). The two types of setup correspond to two different collection points: on the diamond cavity/suspended waveguide (high NA, collection 1) or at the corresponding GC (low NA, collection 2), as illustrated in Fig. 4(d).

To evaluate the cavity performance at cryogenic temperatures, we first characterized the device resonances at 737 nm at the LN GC (collection 2). Fine laser scans were performed across the cavity resonances, and the reflection signal was monitored. The measured spectrum is shown in Fig. 4(g). This device exhibits a quality factor of $5.28 \times 10^4$, is critically coupled to the LN circuit, and shows a resonance dip with 98% contrast. From this, we extract a total scattering-limited Q of $\sim 1.1 \times 10^5$, consistent with state-of-the-art values for thin-film diamond cavities fabricated without the TFLN backbone. This confirms that the integration process preserves the good performance of the diamond cavity, even in a cryogenic environment.

We subsequently perform photoluminescence (PL) characterization of the SiVs implanted in the diamond devices. For this measurement, both collection points are used: we excite with 532 nm CW on the diamond cavity/suspended waveguide, and collect the SiV ZPL at 737 nm, either confocally (collection 1), or at the corresponding GC (collection 2), which is far removed from excitation, as shown in Fig. 4(d).

Emission from an ensemble of SiV can be observed through both collections 1 and 2. Fig. 4(f) shows the spectrum of both, with ABCD peaks corresponding to the SiV ZPLs. These results confirm that SiV emission can successfully propagate through the escalator and LN circuit. Comparing the spectra, we observe that collection 2 through the LN GC shows similar SiV features with broadened lines. There are two possible explanations: (1) Collection 2 likely includes the contribution of many SiVs distributed along the diamond waveguide and taper, whose emission is coupled into the corresponding GC and not detected confocally; (2) we note that the green excitation through the low NA lens is less effective, so the accumulation time for the spectrum is longer (180s instead of 30s) with higher power (100 µW instead of ~10 µW). Based on the C line wavelength, however, this factor is likely less prominent[25].

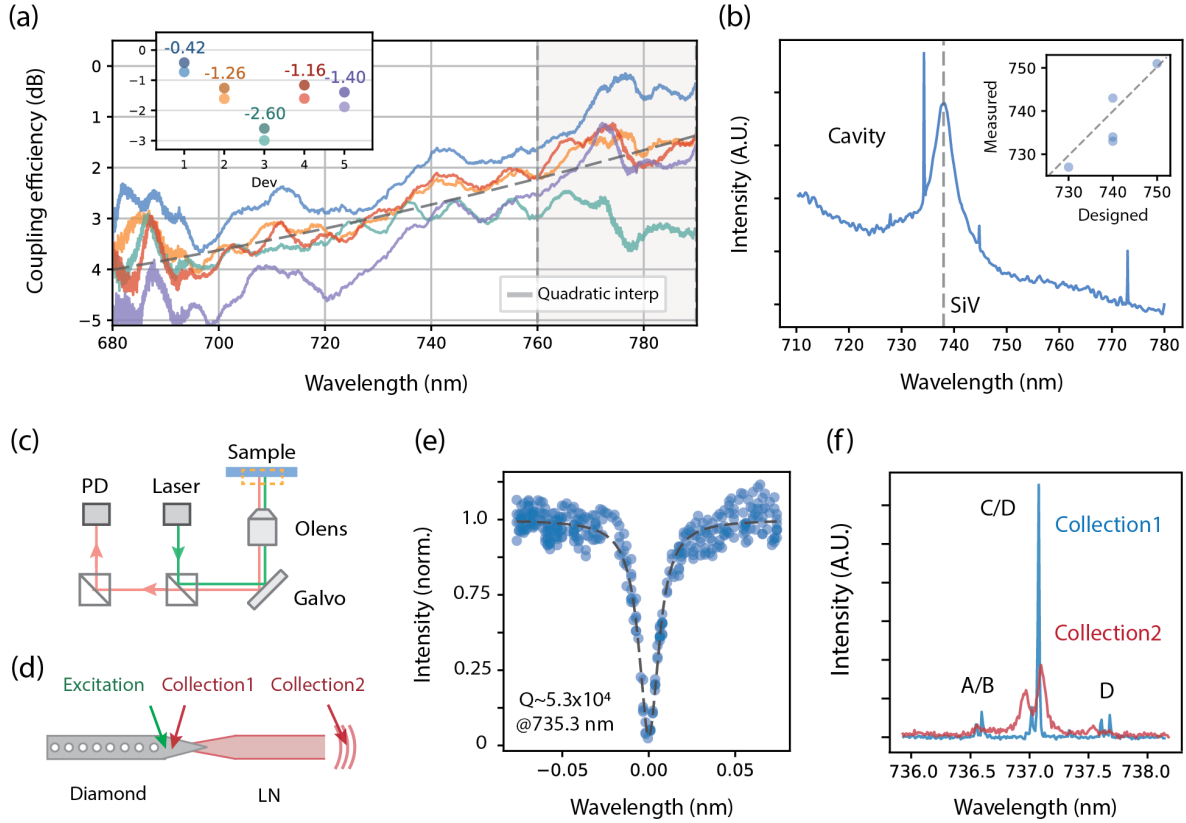

Fig. 4 Characterization of the fabricated sample: (a) The on-chip coupling efficiency ($\eta_{escalator} + \eta_{LN}$) spectrum of the five taper-coupled devices. The efficiency is in dB per facet. The quoted coupling efficiency is for the 760-790 nm range, as indicated by the dashed line and shaded area. Inset: the extracted coupling efficiency of each device is shown with (on-chip, dark) and without (corrected, light) considering the LN propagation loss in the measurements. The labeled values are for the corrected escalator efficiency, with the LN propagation loss accounted for. (b) The measured diamond cavity spectrum of device 1. The dashed line indicates the SiV position at 737 nm, and the fundamental cavity mode is the sharp peak to the left at 735 nm. The inset shows the measured resonance wavelengths compared with the designed values (in nm), with a close match across all five devices. (c) The optical setup for the 5K characterization of the devices. Green (532 nm) is used for excitation. Red (737 nm) is collected via a collection path. Olens: objective lens, 20x or 100x; Galvo: galvo mirror; PD: photodiode. (d) A diagram illustrating the location for excitation and collection on the circuit. (e) The reflection spectra of a cavity at 735 nm, measured by scanning a laser through the resonance using the GC, or collection 2. (f) The spectra of the SiV ZPL signals were collected from collection spots 1 and 2.

## Discussion and conclusion

We present a heterogeneously integrated diamond-on-TFLN platform that unites state-of-the-art diamond photonic crystal cavities containing diamond color centers with thin-film lithium niobate photonic circuits. The integration is accomplished through transfer printing and lithographically controlled alignment. This approach addresses the key figures of merit for hybrid platforms: it is scalable, preserves the material and device quality of both diamond and LN, and enables efficient, mode-matched interfaces. Our LN-diamond escalators achieve a transmission efficiency of up to –0.42 dB/escalator at 775 nm. The cavities exhibit high yield across multiple designs, achieving a Q of $5.3 \times 10^4$ when critically coupled to the circuit, with a scattering-limited Q of $\sim 1.1 \times 10^5$, consistent with the best standalone diamond cavities. At cryogenic temperatures, SiV ensembles are detected through both confocal and on-chip LN collection paths, confirming that the integrated platform is functional at 5K and capable of transmitting SiV emission.

Besides seeing the promise integrating TFLN and diamond and emitters this way, we also note a few directions of future improvements: (1) with larger diamond thin films or multiple transferred, we can achieve more sets of devices in one fabrication run; (2) the adhesion of the diamond to the LN substrate or the transfer process can improved for this process, so that the tapers do not get dislodged and introduce misalignment, and the escalator does not experience the scattering loss related to the transfer residue. This will likely bridge the final 1 dB loss gap and also improve yield; (3) as seen in SiV measurements, sample heating begins to manifest. For our testing of the chip's structural integrity, this was not a problem; however, for future applications involving electrode tuning knobs, heat management is more crucial. It is important to design the circuit and optimize the mounting strategy to alleviate this issue for quantum applications; (4) based on known techniques, such as masked implantation[1], SiV location and density in the diamond can be optimized to couple single centers to the photonic cavities with high cooperativity and with minimal concentration in the waveguide/taper area to reduce fluorescent background.

Overall, the results presented here establish a new fabrication pathway that addresses the limitations of previous hybrid approaches and show an efficient, scalable, and cryo-compatible interface between thin-film diamond and TFLN. It opens the door to large-scale integrated circuits that combine diamond spin qubits with LN's EO and nonlinear capabilities, paving the way for advanced quantum photonic technologies.

## Acknowledgements

This work was supported by the Air Force Lifecycle Management Center under Award No. FA8702-15-D-0001; the Air Force Office of Scientific Research under Award No. FA9550-23-1-0333; DRS Daylight Solutions, Inc. under Award No. A56097; IonQ, Inc. under Award No. A60899; Amazon Web Services under Award No. A60290; and US Department of Energy (DOE) QSA Center (DE-AC02-05CH11231), the National Science Foundation (NSF) (grant PHY-2012023), the Center for Ultracold Atoms (an NSF Physics Frontiers Center), QuEra Computing. Diamond membrane synthesis is primarily funded through Q-NEXT, supported by the U.S. Department of Energy, Office of Science, National Quantum Information

Science Research Centers. Growth-related efforts are also supported by the U.S. Department of Energy, Office of Basic Energy Sciences, Materials Science and Engineering Division. The membrane bonding work is supported by the Quantum Leap Challenge Institute for Hybrid Quantum Architectures and Networks (HQAN) (NSF OMA-2016136), NSF award AM-2240399, and the AFOSR and ONR sponsored MURI on Quantum Phononics (FA9550-23-1-0333). This work made use of the Pritzker Nanofabrication Facility (Soft and Hybrid Nanotechnology Experimental Resource, NSF ECCS-2025633) and the Materials Research Science and Engineering Center (NSF DMR-2011854) at the University of Chicago. The authors thank Dylan Renaud and Michael Haas for valuable discussions and support.